\documentclass[a4paper,USenglish,cleveref]{lipics-v2019}

\usepackage{microtype} % if unwanted, comment out
\usepackage{cite}
\usepackage{subcaption} % subfigures etc.
\usepackage[basic]{complexity}

\graphicspath{{./figures/}}

\usepackage{xcolor}

\title{Recognizing embedded caterpillars with weak unit disk contact representations is \NP-hard%
\footnote{Research partly supported by the German Research Foundation within the collaborative DACH project \emph{Arrangements and Drawings} as DFG Project MU 3501/3-1.}}

\titlerunning{\NP-hardness of caterpillar contact disk representations}

\author{Man-Kwun Chiu}{Institut f\"ur Informatik, Freie Universit\"at Berlin}{chiumk@zedat.fu-berlin.de}{}{Supported by ERC StG 757609.}
\author{Jonas Cleve}{Institut f\"ur Informatik, Freie Universit\"at Berlin}{jonascleve@inf.fu-berlin.de}{https://orcid.org/0000-0001-8480-1726}{Supported by ERC StG 757609.}
\author{Martin N\"ollenburg}{Institute of Logic and Computation, Technische Universit\"at Wien}{noellenburg@ac.tuwien.ac.at}{}{Supported by FWF grant AJS 399.}
\authorrunning{M.-K. Chiu, J. Cleve, and M. N\"ollenburg}
\ccsdesc{Human-centered computing~Graph drawings}
\ccsdesc{Theory of computation~Computational geometry}
\keywords{caterpillar graph, unit disk contact representation, NP-hardness}
\hideLIPIcs{}
\nolinenumbers{}
\begin{document}

\maketitle

\begin{abstract}
	Weak unit disk contact graphs are graphs that admit a representation of the nodes as a collection of internally disjoint unit disks whose boundaries touch if there is an edge between the corresponding nodes.
	We provide a gadget-based reduction to show that recognizing embedded caterpillars that admit a weak unit disk contact representation is \NP-hard.
%	This strengthens a similar result for general embedded trees.
\end{abstract}

\section{Introduction}

A \emph{disk contact graph} $G=(V,E)$ is a graph that has a geometric realization as a collection of internally disjoint disks mapped bijectively to the node set $V$ such that two disks touch if and only if the corresponding nodes are connected by an edge in $E$.
It is well known that the disk contact graphs are exactly the planar graphs~\cite{k-kka-36}.
If, however, all disks must be of the same size, the recognition problem is \NP-hard~\cite{bk-udgrn-98}.
Investigating the precise boundary between hardness and tractability for recognizing unit and weighted disk contact graphs has been the subject of some recent work~\cite{bdlrst-rscplrudct-15,knp-rwdcg-15a,aegkp-bcppg-14,clru-hskudt-16}.
For instance, recognizing embedded trees admitting a unit disk contact representation (UDCR) is \NP-hard~\cite{bdlrst-rscplrudct-15}, while the problem is trivial for paths or stars.
In this paper we study the open problem of recognizing embedded caterpillars that have an embedding-preserving UDCR\@.
% and show that this problem is \NP-hard.

A \emph{caterpillar} $C=(V,E)$ is a tree whose internal nodes form a path, i.e., after removing all leaves from $C$ a \emph{backbone path} remains.
Accordingly we introduce the notions of \emph{leaf} and \emph{backbone} nodes and disks of $C$.
Klemz et al.~\cite{knp-rwdcg-15a} showed that for caterpillars without a given embedding it can be decided in linear time whether a UDCR exists.
Yet, if the cyclic order of the neighbors of each node $v \in V$, i.e., the \emph{embedding} of $C$, is specified and must be preserved, we show that the decision problem is \NP-hard, at least in the following weaker sense.
In a \emph{weak} UDCR of a caterpillar we still require that the disks of any two adjacent nodes in $C$ must touch, yet we also allow that  non-adjacent disks touch.
According to this definition we can obtain dense circle packings on a hexagonal grid with a maximum node degree of $6$, while according to the original definition of (\emph{strong}) UDCRs
%disk contact representations
proper gaps must exist between any pair of non-adjacent nodes and hence if the graph is a tree all nodes have degree at most $5$.
Generalizing the \NP-hardness to strong UDCRs remains an open question.

% We consider the problem of deciding whether a caterpillar graph $G$ with a given embedding can be realized as a unit disk touching graph, i.e., every node represents a unit disk and two disks touch if and only if there is an edge between the corresponding nodes.
% A Caterpillar graph is a tree in which removing all the leaves results in a path graph, which we call the {\em backbone}.
% The given embedding intuitively tells us how many leaves each interior node has to the left and to the right along the backbone.
% %the continuing path of interior nodes.
% In the realization, we call each disk a {\em leaf-disk} or a {\em backbone-disk} according to its role in the caterpillar graph.
% We consider a slightly different setting of the disk touching graph where a leaf-disk is allowed to touch more than one disk, but only one edge will be created between itself and a backbone-disk in the disk touching graph.
% However, any two touching backbone-disks must form an edge.

%We consider a slightly different setting where an edge requires that two disks touch but two disks are also allowed to touch if their corresponding nodes are not neighbors.
%We still do not allow two disks corresponding to two interior nodes to touch.

% We will show that answering this question is \NP-hard.

\section{\NP-hardness Reduction}
\begin{figure}[tbp]
  \centering
  \includegraphics[scale=1,page=1]{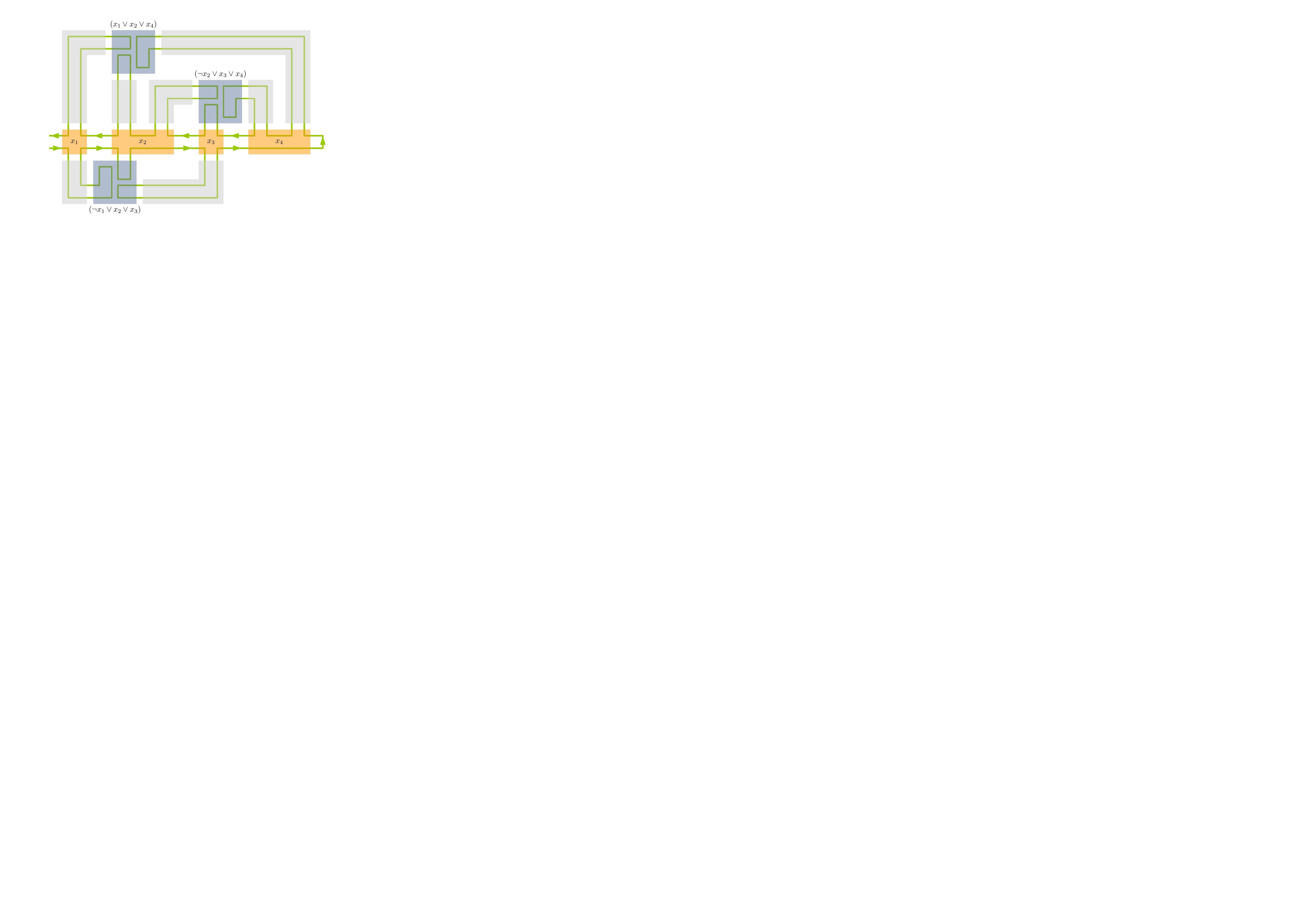}
  \caption{
    A rectilinear drawing of
    %the planar formula
    $(\lnot x_1 \lor x_2 \lor x_3) \land (x_1 \lor x_2 \lor x_4) \land (\lnot x_2 \lor x_3 \lor x_4)$.
    Variables (orange) are connected to their involved clauses (blue).
    % which contain them.
    The caterpillar will follow the green path.
  }\label{fig:planar-3sat}
\end{figure}
To prove our \NP-hardness result, we reduce from the \NP-complete problem \textsc{Planar 3-SAT}.
%Before sketching the detailed reduction, we explain a birds-eye view.
We first give an overview and then describe the gadgets in detail.
Given a \textsc{Planar 3-SAT} instance $\phi$ with $n$ variables and $m$ clauses and its planar variable-clause graph $G(\phi)$, we construct an embedded caterpillar of size $O(m^2+nm)$ that admits a weak UDCR if and only if $\phi$ is satisfiable.
First, we will design a caterpillar $C$ with a unique high-level realization that mimics the planar drawing of the variable-clause graph $G(\phi)$ of the \textsc{Planar 3-SAT} formula $\phi$ (see \cref{fig:planar-3sat}).
The unique realization can be obtained by locally optimal packings of leaf disks to enforce the required grid positioning of the backbone disks (see \cref{fig:rigidity:fixes-path}).
We also call this a rigid construction.
We then modify the subgraph of $C$ near each variable in $G(\phi)$ such that we now have two possible local realizations corresponding to the true/false assignments in each variable gadget.
The position of the realization will be propagated through the rigid components to the involved clause gadgets such that each clause gadget can be realized if and only if at least one of its literals is true.
We obtain
%With this we will prove the following
\begin{theorem}\label{theorem}
  The problem of deciding whether a caterpillar with a given embedding admits a weak unit disk contact representation in the plane is \NP-hard.
\end{theorem}

% We further note that for unit disk realizations of caterpillars an embedding actually only defines for each backbone node how many of its neighboring leaves lie on either side of the backbone path.

%The base idea is to force the realization consisting of several rigid components connected by some joints, which only allow limited choices in the realization. The rigid components mean that there is only one realization of the corresponding subgraph.
%The limited freedom created by the joints would represent the true/false assignment

\subsection{Planar 3-SAT}

Given a Boolean formula $\phi$ in 3-CNF with $n$ variables, its corresponding variable-clause graph $G(\phi)$ has nodes for each variable $x_i$ and each clause $c_j$ of $\phi$ and there are edges between a variable $x_i$ and a clause $c_j$ iff either $x_i$ or $\lnot x_i$ appear in $c_j$.
Furthermore there is an edge $\{x_i,x_{i+1}\}$ for all $1\leq i<n$ plus $\{x_n,x_1\}$, i.e., a cycle through all variable nodes.
The \textsc{Planar 3-SAT} problem is to decide, given a formula $\phi$ for which $G(\phi)$ is planar, whether $\phi$ is satisfiable.
Lichtenstein~\cite{Lichtenstein:Planar_formulae_their:82} showed that \textsc{Planar 3-SAT} is \NP-complete.
%Due to the variable cycle i
It is possible to arrange all variables on a horizontal line and to use only rectilinear connectors to connect the variables with the respective clauses in a comb-like fashion~\cite{Knuth:problem_compatible_representatives:92}. %, called \textsc{Planar Rectilinear 3-SAT}.
An example is shown in \cref{fig:planar-3sat} where we added a directed path indicating how the caterpillar traverses $G(\phi)$.

\subsection{Rigidity -- Allowing Exactly One Realization (up to Rotation)}

\begin{figure}[tbp]
  \begin{subfigure}[t]{0.28\linewidth}
    \centering
    \includegraphics[scale=1.3,page=1]{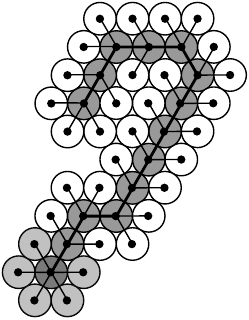}
    \caption{
      By starting with an interior node with 5 leaves this is the graph's only realization (up to rotation).
    }\label{fig:rigidity:fixes-path}
  \end{subfigure}
  \hfill
  \begin{subfigure}[t]{0.65\linewidth}
    \centering
    \includegraphics[scale=1.3,page=2]{rigidity}
    \caption{
      Five (of infinitely many) different realizations when allowing freedom after the second interior node.
      Nodes which can move around are marked.
      The third and fourth interior nodes together regain rigidity.
      The fourth disk stays in the marked $\pi/3$ sector.
    }\label{fig:rigidity:can-be-regained}
  \end{subfigure}

  \caption{
    An interior node with six neighbors enforces rigidity, even after allowing some freedom.
  }\label{fig:rigidity}
\end{figure}

We first observe that we can use a locally optimal packing of unit disks to enforce the direction in which the caterpillar continues.
As observed in \cref{fig:rigidity:fixes-path}, starting with a node with five leaves fixes the position of the next backbone-disk.
Since the next backbone-disk can have up to 3 more neighbors, adding two leaves to this node in a particular cyclic order again fixes the next backbone-disk's position.
By repeatedly applying this construction, we can build a rigid 3-disk-wide path on a hexagonal grid.
%With a construction like this we can trace arbitrary 3-disk-wide paths on a hexagonal grid.
Furthermore, as shown in \cref{fig:rigidity:can-be-regained}, even after allowing some freedom of movement, rigidity can be regained by an interior node with four leaves and thus six neighbors.
Hence, it is possible to have two rigid components joined by a non-rigid part.
\begin{figure}[tbp]
  \centering
  \includegraphics[scale=1.3,page=1]{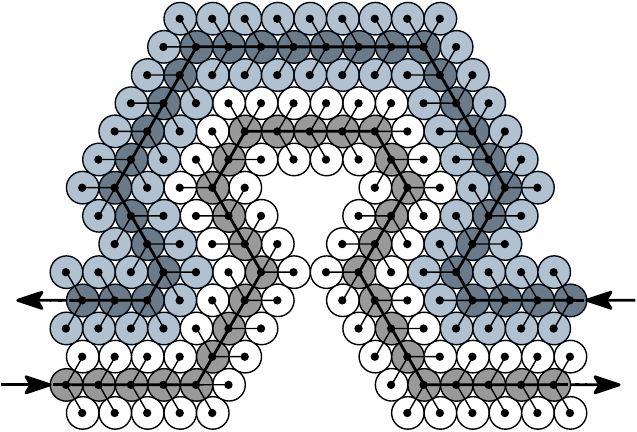}
  \caption{
    If the lower and upper part travel in the annotated direction and cannot move vertically apart by more than six disks, this is the only possible realization.
  }\label{fig:two-path-lock}
\end{figure}
Sometimes we would like to make sure that two parts of the caterpillar have a certain position relative to each other.
This can be achieved by introducing a locking structure which is shown in \cref{fig:two-path-lock}.

\subsection{Variable Gadgets -- Allowing Exactly Two Different Realizations}

\begin{figure}[tbp]
  \begin{subfigure}[t]{\linewidth}
    \centering
    \includegraphics[scale=1.5,page=1]{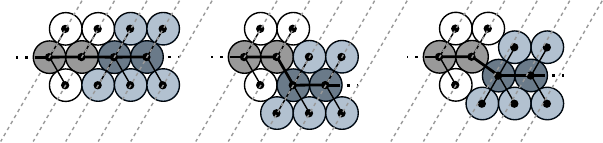}
    \caption{
      Reparenting one leaf to the next interior node gives a restrained freedom of movement.
      All positions between the upper (left) and lower (center) position are possible (right).
    }\label{fig:two-values:continuous}
  \end{subfigure}
  \par\vspace{1em}
  \begin{subfigure}[t]{\linewidth}
    \centering
    \includegraphics[width=\linewidth,page=3]{two-values}
    \caption{
      Adding a rigid structure which is aligned with the hexagonal grid removes the possibility to realize the intermediate positions (right).
      Only the two extremal positions (left and center) remain.
    }\label{fig:two-values:discrete}
  \end{subfigure}

  \caption{
    A construction which allows for exactly two different realizations.
  }\label{fig:two-values}
\end{figure}

For the reduction we design a caterpillar and its embedding in such a way that there are exactly two local realizations to simulate truth values in each variable gadget.
As shown in \cref{fig:two-values:continuous}, flipping the connection of one leaf to the next interior node along the rigid path allows the latter path to shift between two positions where the line passing through the two positions forms an angle of 60 degrees relative to the direction of the backbone.
%For the reduction we want to give a caterpillar with an embedding which allows for more than one realization to simulate truth values.
%As can be seen in \cref{fig:two-values:continuous}, moving one leave to the next interior node in an otherwise rigid path allows the second part of the path to shift between two positions which are aligned with the hexagonal grid.
However, intermediate positions are also possible which we want to prevent.
Since the movement happens along a circular arc all intermediate positions might cause an intersection in other grid-aligned paths.
By deliberately introducing such a path, as shown in \cref{fig:two-values:discrete}, we can restrict this part of the graph to be realized in only two possible ways.

\begin{figure}[tb]
  \hfill
  \includegraphics[width=0.25\linewidth,page=1]{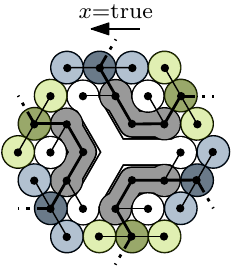}
  \hfill
  \includegraphics[width=0.25\linewidth,page=2]{variable-concept}
  \hfill
  \includegraphics[width=0.25\linewidth,page=3]{variable-concept}
  \hfill\null% \null is important here, otherwise \hfill is ignored
  \caption{
    The variable gadget idea.
    Assume that the white-gray hexagon in the center is somehow fixed.
    Then, moving one of the colored parts in one direction forces the movement of all five others.
    Again, we have two extremal positions (left and center) but also all intermediate positions (right).
  }\label{fig:variable-concept}
\end{figure}

\begin{figure}[tb]
  \hfill
  \includegraphics[width=0.32\linewidth,page=1]{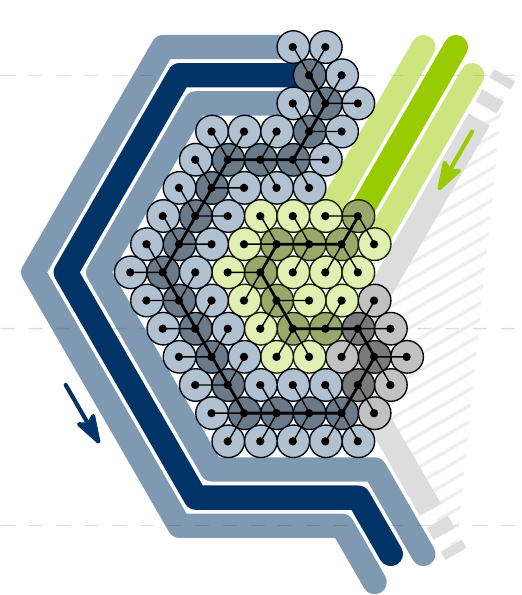}
  % \hfill
  \includegraphics[width=0.32\linewidth,page=2]{variable-corner}
  % \hfill
  \includegraphics[width=0.32\linewidth,page=3]{variable-corner}
  \hfill\null% \null is important here, otherwise \hfill is ignored
  \caption{
    Introducing an interlocking structure at the corner of the variable gadget prohibits all but the two extremal positions (left and center).
    It also prevents any movement of the gray part.
  }\label{fig:variable-corner}
\end{figure}

\begin{figure}[tb]
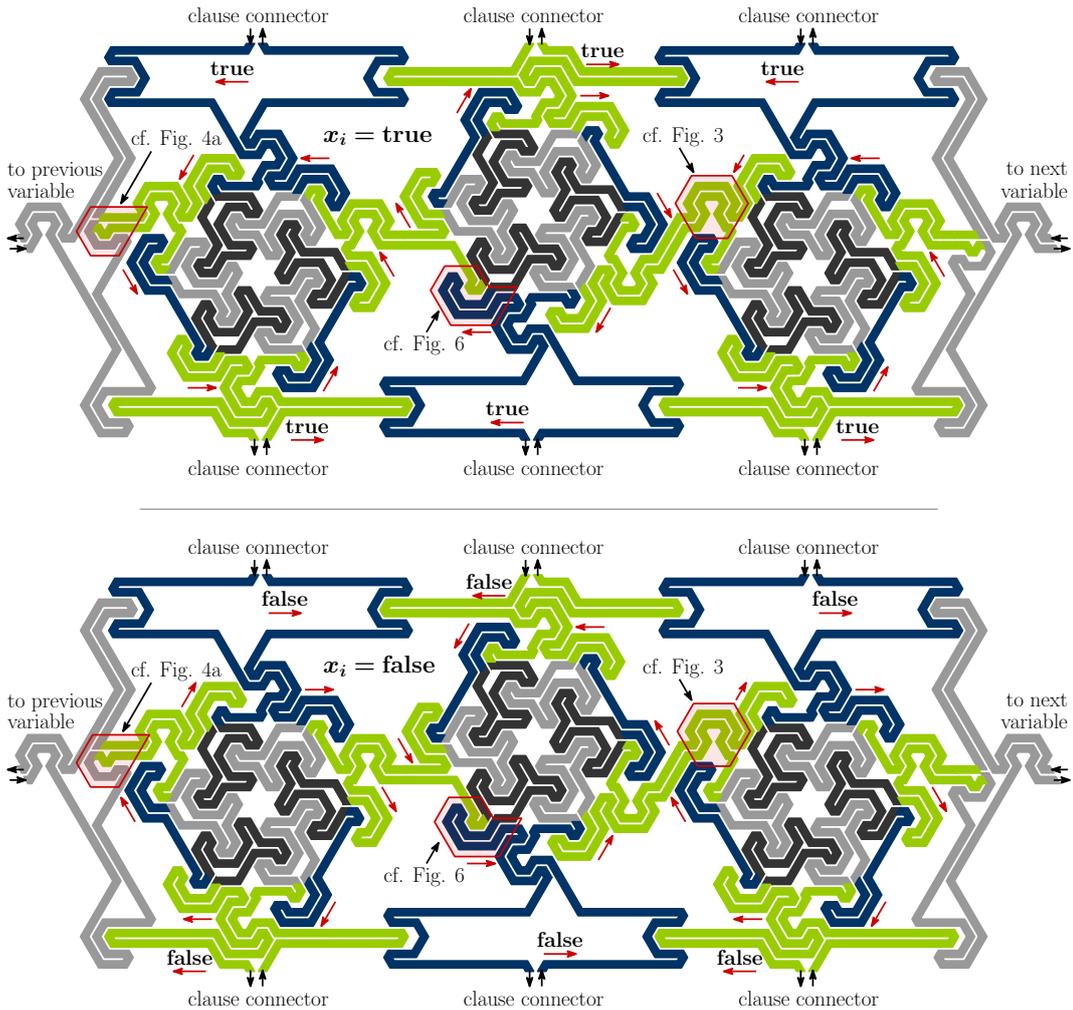

  \centering
  \includegraphics[width=\linewidth,page=4]{variable-gadget}
  \vspace{1em}
  \textcolor{gray}{\rule{.75\linewidth}{0.4pt}}
  \vspace{1em}
  \includegraphics[width=\linewidth,page=5]{variable-gadget}
  \caption{
    A simplified variable gadget depiction for $x_i=\text{true}$ (top) and $x_i=\text{false}$ (bottom).
    All six clause connectors are shifted by one disk to the left or right compared to the other state, indicated by the red arrows.
    Repeating the last two hexagons adds more connectors.
    Chaining the whole gadget gives arbitrarily many variables.
    Some appearances of previous figures are highlighted.
  }\label{fig:variable-gadget}
\end{figure}

In \cref{fig:variable-concept} we show the basic idea of the variable gadget.
We assume to have a fixed inner structure (represented by the uncolored inner hexagon) to which six different paths are connected.
Those paths are colored in alternating colors to distinguish them easily.
The outer paths can be pushed in a counter-clockwise or clockwise fashion (which can be interpreted as $x=\text{true}$ or $x=\text{false}$) which moves exactly one disk in each of the six main directions.
However, as before, intermediate positions are possible but they have to be avoided.
By making the hexagon bigger, we can use a similar construction as before to only allow the two extremal positions in any realization of the caterpillar.
The solution to this can be found in \cref{fig:variable-corner} which focuses on just one corner with two adjacent paths (a sketch of the full hexagon can be seen in \cref{fig:variable-gadget}).
The gray part to the right is a corner of the inner hexagon and considered fixed. %\kenny{fixed, really? It seems \cref{fig:variable-corner} not matching \cref{fig:variable-gadget}.}
If the green path is pushed to one extremal position, the blue path has to follow so that no overlapping occurs.
If the green path is in an intermediate position, the blue hook cannot align itself with the green path without intersection such that it is still touching the gray disk following the caterpillar.

All these ideas are combined to form a full variable gadget (see \cref{fig:variable-gadget}).
%which is shown in \cref{fig:variable-gadget} in the two possible realizations.
The caterpillar path is assumed to be rigid when entering the gadget from the left.
Each part with the same color is completely rigid and the transitions between two colors are as in \cref{fig:two-values} so that we only have two possible local realizations.
The path first traces the gray part on the lower left which is to prevent movement of the hexagons in the up-down-direction.
Afterwards the construction from \cref{fig:two-values:discrete} is used to allow exactly two positions for the following part.
The lock from \cref{fig:two-path-lock} will make sure that the corresponding part on its way back will be together with the current part.
The path moves counter-clockwise around the hexagon while using the construction from \cref{fig:variable-corner} for the corners and simultaneously some interlocking path for the inner hexagon to make the interior completely rigid.
When reaching the bottom part of the outer hexagon we extend to the left to align the vertical position of the variable with the outer structure.
Then the path goes towards a clause and comes back to the same place---if no clause is connected here, we just connect the two parts directly.

We continue on the lower side of the construction into the next hexagon.
%Note that i
If the first hexagon is pushed clockwise the second one is pushed counter-clockwise and vice-versa.
This means, that every second clause connector is pushed left while every other second is pushed right.
We finally finish the lower part of the construction of one variable gadget by reaching the gray part on the lower right.
Here the path enters another variable gadget and eventually comes back to trace the upper part of the construction in the same fashion as the lower part.
To have more than six clause connections shown here, we can just repeat the first and second hexagons arbitrarily often.
Different variables are just chained to the right (cf. \cref{fig:planar-3sat}).

\subsection{Clause Gadgets}

\begin{figure}[tbp]
  \hfill
  \includegraphics[width=0.24\linewidth,page=1]{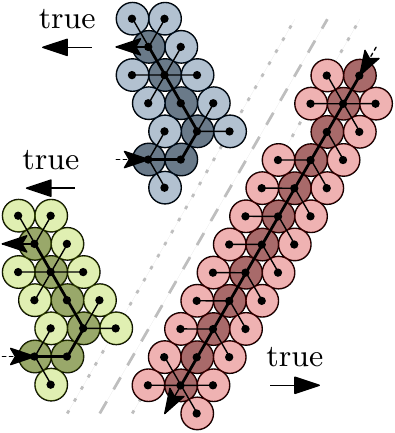}
  \hfill
  \includegraphics[width=0.24\linewidth,page=2]{clause-concept}
  \hfill
  \includegraphics[width=0.24\linewidth,page=3]{clause-concept}
  \hfill
  \includegraphics[width=0.24\linewidth,page=4]{clause-concept}
  \hfill\null% \null is important here, otherwise \hfill is ignored
  \caption{
    The idea of the clause gadget: The two literals on the left have one bulge each whereas the literal on the right has one notch which can accomodate either but not both bulges.
  }\label{fig:clause-concept}
\end{figure}

We now have a variable gadget which moves a rigid sub-caterpillar between exactly two possible positions on the hexagonal grid, namely left and right.
With this we want to construct a clause gadget which should be realizable if and only if at least one literal is set to true.
The idea for the clause gadget is shown in \cref{fig:clause-concept}:
We have one larger part coming from the right which has exactly one leaf missing and two smaller parts coming from the left, each of which has one leaf protruding to the right.
The three parts should be connected to the corresponding variable gadgets such that a true value for the corresponding literal pulls them away from the center and a false value pushes them towards the center.
As we can observe, if the right part is set to false there is room for at most one leaf of a left part but not for both.
Thus, setting all literals to false makes it impossible to realize this caterpillar, while in all other cases a realization of the clause gadget exists.

\begin{figure}[tbp]
  \centering
  \includegraphics[width=0.8\linewidth,page=1]{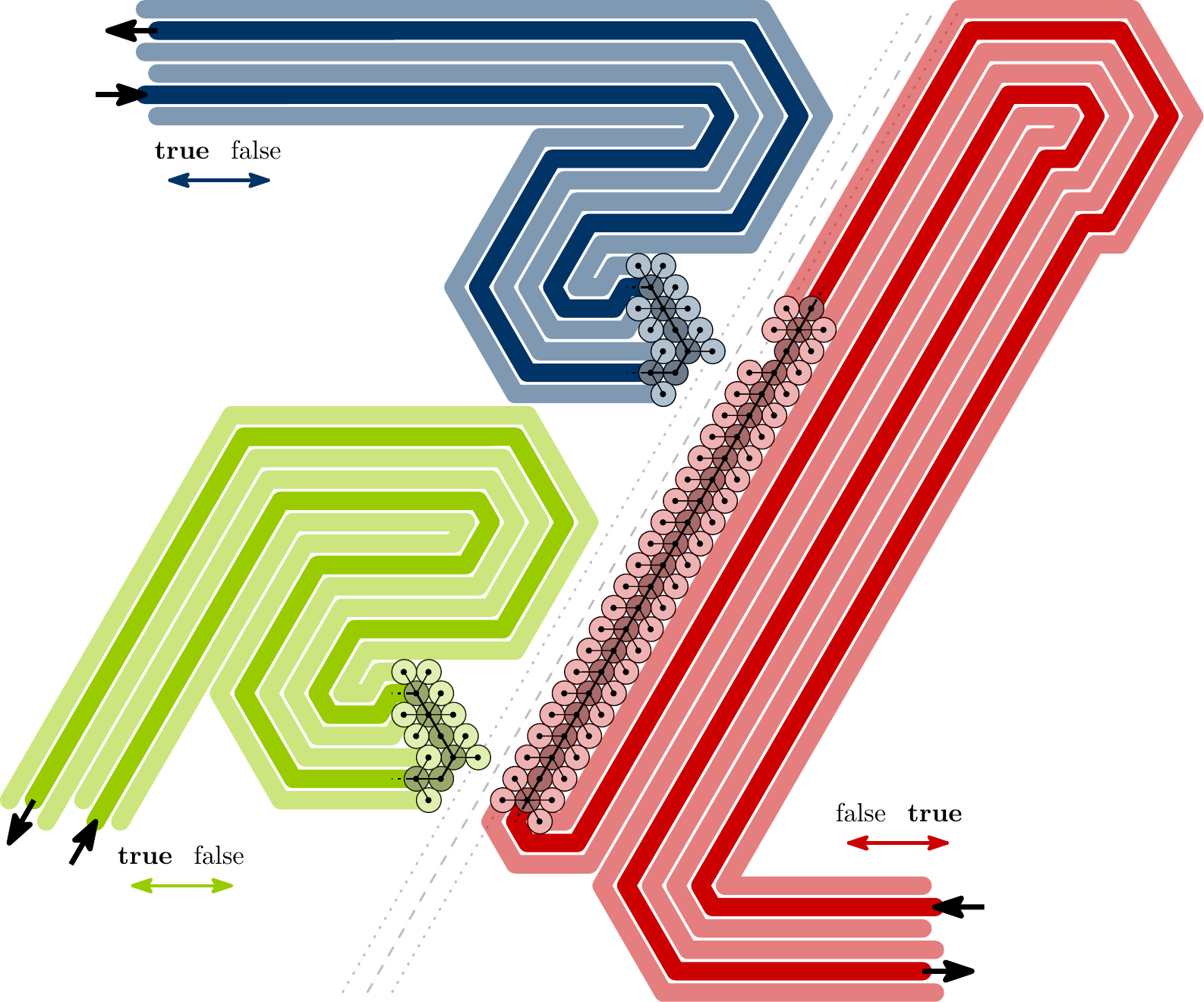}
  \caption{
    The full clause gadget.
  }\label{fig:clause-gadget}
\end{figure}

\begin{figure}[p]
  \centering
  \includegraphics[width=.9295\textheight,page=1,angle=90]{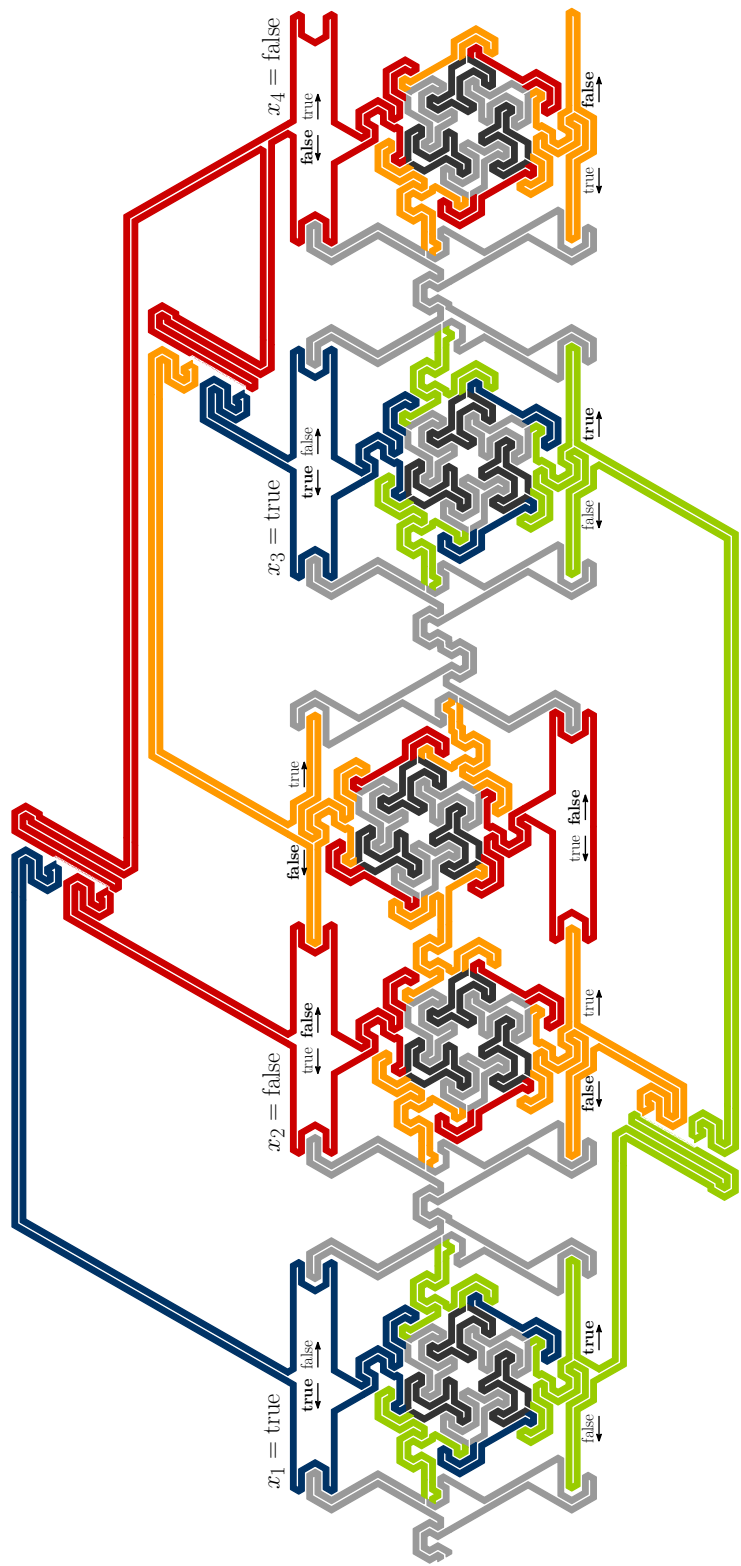}
  \caption{
    One possible realization of the formula from \cref{fig:planar-3sat} with $x_1=\text{true}$, $x_2=\text{false}$, $x_3=\text{true}$, and $x_4=\text{false}$.
    Due to space constraints the first hexagon of $x_4$ behaves different from the other variables.
    Otherwise a second hexagon would be needed.
  }\label{fig:full-picture-caterpillar}
\end{figure}

Each of the three parts has some missing leaves and thus causes some freedom to move around.
We need to make sure that, despite possible movement, they can be only realized the way we intend.
The result is shown in \cref{fig:clause-gadget}.
The long right side of the clause will be realized as the first part---because of the one missing leaf it could be rotated by at most 60 degrees.
By going all the way back with a small interlocking on the top this would lead to self-intersection and thus the shown realization is the only one (ignoring the leaves which could move up and down).
The two paths on the left can only be realized as shown because they would otherwise intersect with the right side or with themselves.
The clause gadget is connected like this on the upper side of the whole construction and rotated by 180 degrees on the lower side of the construction.
We finally show a full picture of one possible realization of the abstract drawing of \cref{fig:planar-3sat} in \cref{fig:full-picture-caterpillar}.

\subsection{Summary of the Reduction}
Each variable gadget starts and end with a rigid part with constant size.
Furthermore, each literal needs at most two hexagons (of constant size) in its corresponding variable gadget to have the correct connector.
We have exactly $3m$ literals in $\phi$ and hence we need $O(n+m)$ many nodes for the variable gadgets.
Each clause gadget has constant size and sits on its individual level.
We can have at most $m$ levels and each of the three connectors per clause has height and width of at most $O(m)$ and $O(n+m)$ respectively.
Thus the clause gadgets with the connectors need $O(mn + m^2)$ many nodes which is also the size of the full construction.

Since the variable gadgets always start the same way, a variable is set to true if and only if the first hexagon is rotated counter-clockwise, false otherwise.
Hence, by design of the gadgets above, a formula $\phi$ is satisfiable if and only if the corresponding polynomial-size caterpillar $C(\phi)$ can be recognized as a weak UDCR\@.
This concludes the proof of \cref{theorem}.

\subparagraph*{Acknowledgments.} This work was initiated during the
\emph{Japan-Austria Bilateral Seminar: Computational Geometry Seminar with Applications to Sensor Networks} in Zao Onsen, Japan in November 2018.
We thank the organizers for providing a productive environment and the other participants, especially Oswin Aichholzer, André van Renssen, and Birgit Vogtenhuber, for the initial discussions.

\bibliography{caterpillar}

\end{document}